\documentclass[twocolumn,pre,superscriptaddress]{revtex4-2}
\usepackage[utf8]{inputenc}
\usepackage{times}
\usepackage[dvips]{graphicx}
\usepackage[usenames,dvipsnames]{xcolor}
\usepackage{amsmath}
\usepackage{amssymb}
\usepackage{bm}

\newcommand{\figref}[1]{{\color{orange}\ref{#1}}}
\newcommand{\kBT}{k_{\rm B}T}

\newcommand{\Revised}[1]{{\color{black}#1}}

\begin{document}

\title{Increase in Rod Diffusivity Emerges even in Markovian Nature}

\author{Fumiaki Nakai}
\email{nakai.fumiaki.c7@s.mail.nagoya-u.ac.jp}
\affiliation{Department of Materials Physics, Graduate School of Engineering, Nagoya University, Furo-cho, Chikusa, Nagoya 464-8603, Japan}

\author{Martin Kr\"oger}
\email{mk@mat.ethz.ch} 
\affiliation{Polymer Physics, Department of Materials, ETH Zurich, CH-8093 Zurich, Switzerland}
\affiliation{Magnetism and Interface Physics, Department of Materials, ETH Zurich, CH-8093 Zurich, Switzerland}

\author{Takato Ishida} 
%\email{} 
\affiliation{Department of Materials Physics, Graduate School of Engineering, Nagoya University, Furo-cho, Chikusa, Nagoya 464-8603, Japan}

\author{Takashi Uneyama}
%\email{} 
\affiliation{Department of Materials Physics, Graduate School of Engineering, Nagoya University, Furo-cho, Chikusa, Nagoya 464-8603, Japan}

\author{Yuya Doi}
%\email{} 
\affiliation{Department of Materials Physics, Graduate School of Engineering, Nagoya University, Furo-cho, Chikusa, Nagoya 464-8603, Japan}

\author{Yuichi Masubuchi}
%\email{} 
\affiliation{Department of Materials Physics, Graduate School of Engineering, Nagoya University, Furo-cho, Chikusa, Nagoya 464-8603, Japan}

% \date{March 2022}

\begin{abstract}
% Rod-shaped particles embedded in certain matrices have been reported to exhibit an increase in their center of mass diffusivity upon increasing the matrix density.
% This increase has been considered to be caused by a kinetic constraint in analogy with tube models.
% Here, we investigate a mobile rod-like particle in a three-dimensional sea of immobile point obstacles using a kinetic Monte Carlo scheme equipped with a Markovian process, that generates gas-like collision times and positions stochastically, so that such kinetic constraints do essentially not exist.
% We find that even in such a system, the unusual increase in diffusivity emerges.
% This result implies that the kinetic constraint is not a necessary condition for the increase in the diffusivity.
%More generally, this work will provide fresh insight into the kinetics of non-spherical particles.

Rod-shaped particles embedded in certain matrices have been reported to exhibit an increase in their center of mass diffusivity upon increasing the matrix density.
This increase has been considered to be caused by a kinetic constraint in analogy with tube models.
We investigate a mobile rod-like particle in a sea of immobile point obstacles using a kinetic Monte Carlo scheme equipped with a Markovian process, that generates gas-like collision statistics, so that such kinetic constraints do essentially not exist.
Even in such a system, \Revised{provided the particle's aspect ratio exceeds a threshold value of about 24,} the unusual increase in the rod diffusivity emerges.
This result implies that the kinetic constraint is not a necessary condition for the increase in the diffusivity.
\end{abstract}
\maketitle

\section{introduction}
The translational diffusion coefficient $D_c$ of a particle is generally known to decrease with increasing matrix density or increasing amount of obstacles.
It is understood as a consequence of the ballistic particle motion being disturbed during collisions with the surrounding matrix.
However, if the particle is rod-shaped, a counter-intuitive motion can occur; the $D_c$ of a rod may increase as the matrix concentration increases, provided the concentration is sufficiently high.
Frenkel and Maguire \cite{frenkel1981molecular, frenkel1983molecular} first observed such behavior for fluids consisting of infinitely thin rods, whose static properties are exactly the same as those of an ideal gas. This finding was later confirmed with higher accuracy \cite{magda1986transport, magda1988transport}.
Their systems do not have hidden particles or thermostats; the constituent particle moves ballistically between elastic collisions.
Following the previous studies\cite{frenkel1981molecular, frenkel1983molecular}, an increase in $D_c$ has been observed in various systems: (i) an infinitely thin rod in a two-dimensional (2D) sea of fixed point obstacles \cite{hofling2008enhanced}, (ii) a thick rod in a 2D matrix of circular obstacles \cite{tucker2010observation}, and (iii) an active matter fluid consisting of rods swimming in direction of its major axis \cite{mandal2020crowding}.
In these systems, the increase in $D_c$ is not triggered by a phase transition.
Still, some rod systems exhibit an increase in $D_c$ accompanied by the isotropic-nematic transition \cite{allen1990diffusion}. Such multi-particle effects remain beyond the scope of the present work.

Various loosely defined concepts have been considered previously to explain the increase in $D_c$: so-called dynamic correlation, steric hindrance, confinement, or tube \cite{frenkel1983molecular, hofling2008enhanced}\Revised{, which represent an assumed cylindrical constraint and disturbs the rod rotation}.
We refer to these concepts as the "kinetic constraint" in what follows.
In this work, we define the kinetic constraint as the constraint that prevents the rod from crossing an obstacle until the rod moves about the rod length. Using the kinetic constraint, the increase in $D_c$ can be explained.
Namely, the rotational motion of the rod is kinetically constrained via the surrounding matrix in the concentrated matrix regime.
Even in such a regime, the ballistic motion along the major axis of an infinitely thin rod is not hindered, while the relevance of collisions in direction of the major axis increases with increasing width of the rod or size of the obstacles.
Consequently, the ballistic motion with the major axis may persist for a relatively long time. This duration may increase with matrix density or the degree of confinement and ultimately leads to an increase in $D_c$.
In the so-called active rod fluid \cite{mandal2020crowding}, a similar behavior is caused by swimming along the axial direction of the rod instead of ballistic motion.
In light of these studies one question may arise; Is the kinetic constraint a necessary condition for the emergence of the increase in diffusivity?

On our way towards an answer, we have been guided by our naive belief that such an increase can be caused by the reduction of the rotational diffusivity alone, without the hindrance of the axially directed motion.
To test our hypothesis rigorously, we consider a simple model system where the rotational diffusivity reduces with increasing matrix density, whereas the ballistic motion along the major axis of the rod-like particle remains largely undisturbed.
One possible such system is a single mobile rod embedded in a three-dimensional (3D) arrangement of spatially fixed point obstacles.
It can be regarded as the extension of the Lorentz gas systems \cite{lorentz1905motion, alder1983decay, hofling2007crossover}; a single spherical particle in fixed obstacles.

In this work, we report that the upturn of $D_c$ emerges even in the presence of a Markovian process where the kinetic constraint does essentially not exist%
% \Revised{;
% the obstacles do not keep in the same place nor limit \mkcomment{(but still affect?)} the rotational motion of the rod.} \mkcomment{add somewhere: as opposed to a standard implementation}
\Revised{;
although the obstacles change the rod motion, they do not keep in the same place nor constrain the rotational motion of the rod.}
% We investigate the trajectories of a sphero-cylinder in a 3D matrix of stochastically homogeneously distributed point obstacles using a Markovian kinetic Monte Carlo (KMC) scheme \cite{gillespie1976general, bortz1975new}.
\Revised{We investigate the trajectories of a sphero-cylinder in a 3D matrix of statistically homogeneously distributed point obstacles using a kinetic Monte Carlo (KMC) scheme \cite{gillespie1976general, bortz1975new}. In this calculation, we assume the Markovian process, where the sequential collisions between the rod and obstacles are uncorrelated like for a dilute gas system, as opposed to a standard implementation of the molecular dynamics simulation.}
The $D_c$ of this rod-like particle increases in an intermediate matrix density regime if the rod is sufficiently long.
In our system, $D_c$ reaches a peak value and subsequently decreases with increasing obstacle density due to the thickness of the rod.
On the basis of the Markovian nature, we give scaling relations between $D_c$ and the obstacle density for dilute, intermediate, and concentrated density regimes.
This work will generate fresh insight into the kinetics of the non-spherical shaped particles.

\section{Model and methods}

\begin{figure}
    \centering
    \includegraphics[width=0.7\columnwidth]{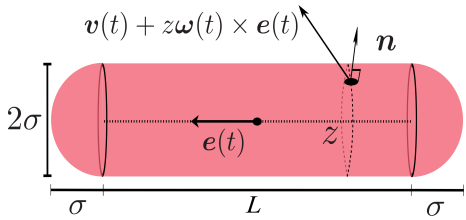}
    \caption{Schematic figure of the rod. The collision frequency for a given $\bm{e}(t)$, $\bm{v}(t)$, and $\bm{\omega}(t)$ is characterized by $z$ and $\bm{n}$.}
    \label{fig:rod_variables}
\end{figure}

The model consists of a rod-like sphero-cylinder (also termed capsule or stadium of revolution) with radius $\sigma$, mass $M$, and length $L$ of its major axis.
The effective "rod" length is $L_e=L+2\sigma$ due to the half-spherical end-caps,
and the inertia tensor $\bm{I}$ is determined by assuming that the mass is homogeneously distributed over the volume of the rod \cite{pournin2005three}.
\Revised{The schematic figure of the rod is displayed in Fig.\ \figref{fig:rod_variables}.}
The point obstacles are statistically homogeneously distributed in the unbounded 3D space at number density $\rho$.
The interaction between the rod and obstacles is modeled by a hard-core potential; the obstacles do not penetrate the rod, and they do not move during a collision.
The rod ballistically moves except when it elastically collides with an obstacle.
% The center of mass velocity $\bm{v}$ and angular velocity $\bm{\omega}$ change during a collision, conserving the rod particle's translational and rotational kinetic energy.
% The total energy of this system is $5\kBT/2$, where $k_{\rm B}$ and $T$ are the Boltzmann constant and temperature.
% Due to the elastic collisions, the total energy is a conserved quantity and does not change during the course of time. 
\Revised{The center of mass velocity $\bm{v}$ and angular velocity $\bm{\omega}$ change during a collision by conserving the total energy of the system, which is distributed over the rod particle's translational and rotational kinetic energy.
The total energy of this system is set to be $5\kBT/2$, where $k_{\rm B}$ and $T$ are the Boltzmann constant and temperature, and does not change during the course of time.}
\Revised{If we choose $M$, $\sigma$, and $\kBT$ to define dimensionless units, the remaining parameters are only the effective rod length $L_e$ and the number density of the obstacles, $\rho$.
For the convenience of the presentation, we define the speed unit $u=\sqrt{\kBT/M}$.
This work displays the physical quantities with dimensions for physical clarity.
If one prefers to work with 
reduced quantities, all the variables $M$, $\sigma$, $\kBT$, and $u$ can be set to unity without loss of generality.
Since the interaction between the rod and obstacles is the hard-core potential, the dynamical properties are independent of the temperature when the physical quantities are expressed using dimensionless units.
}
Here, we \Revised{do not study the trapping regime, which occurs at densities exceeding the inverse volume of the sphero-cylinder, $\rho \sigma^2L_e \gtrsim 1$}.

\begin{figure}[tbp]
\centering
  \includegraphics[width=\columnwidth]{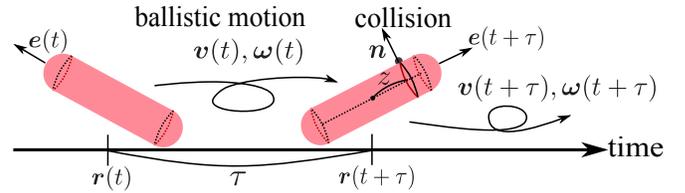}
\caption{Schematic representation of the KMC method. $\bm{r}(t)$, $\bm{e}(t)$, $\bm{v}(t)$, and $\bm{\omega}(t)$ are the position, direction unit vector, velocity, and angular velocity, respectively, of the rod at time $t$. $z$ and $\bm{n}$ characterize the coordinate of the collision point; $z\in[-L/2,L/2]$ is the axial coordinate and $\bm{n}$ is the surface normal at the collision point. $\tau$ is the collision time interval between successive collisions.
$\tau$, $z$, and $\bm{n}$ are stochastically sampled from the collision statistics, Eq.~\eqref{eq_frequency}. From the sampled variables, $\bm{r}$, $\bm{e}$, $\bm{v}$ and $\bm{\omega}$ at time $t+\tau$ are obtained.}
\label{fig_system_kmc_variables}
\end{figure}

To calculate the dynamics of a rod subject to a Markovian collision process, we extend the kinetic Monte Carlo (KMC) simulation method \cite{gillespie1976general,bortz1975new}.
It requires two inputs; (i) statistics of collisions and (ii) the change of dynamical variables by a collision.
For (i), we here extend the calculations for a sphere \cite{dorfman2021, mazenko2008, nakai2023fluctuating} to the collision statistics of a sphero-cylinder and obtain the collision frequency with the coordinates of collision \Revised{at time $t$} for a given \Revised{velocity} $\bm{v}(t)$, \Revised{angular velocity} $\bm{\omega}(t)$, and the direction vector of the rod $\bm{e}(t)$.
In the following, we denote $\bm{\Gamma}(t)$ as the 8-dimensional time-dependent phase space variable $({\bf v}(t), \bm{\omega}(t),{\bf e}(t))$ characterizing
the state of the rod (Fig.~\figref{fig_system_kmc_variables}).
The explicit expression for the collision frequency for a given $\bm{\Gamma}(t)$, $F(\bm{\Gamma}(t))$, arises from a surface integral of the collision frequency density
\begin{eqnarray}
%\begin{split}
    f(z,\bm{n}; \bm{\Gamma}(t)) 
&=&\rho\Revised{\sigma}\bm{v}_e(z;\bm{\Gamma}(t))\cdot\bm{n}
\Theta[\bm{v}_e(z;\bm{\Gamma}(t))\cdot\bm{n}]\nonumber \\
&&\times\left\{
\delta(\bm{e}(t)\cdot\bm{n})
+\Revised{\sigma}\delta\left(z-L/2\right)\Theta[\bm{e}(t)\cdot\bm{n}] \right. \nonumber \\
&&\left.\quad +\Revised{\sigma}\delta\left(z+L/2\right)\Theta[-\bm{e}(t)\cdot\bm{n}]
\right\}
%\end{split}
\label{eq_frequency}
\end{eqnarray}
where $z$ is the axial coordinate along the rod direction and $\bm{n}$ a unit vector normal to the rod's surface (Fig.\ \figref{fig_system_kmc_variables}). These two variables characterize the coordinate $z\bm{e}+\bm{n}$ of the collision point between the rod and an obstacle, while $\bm{v}_e(z;\bm{\Gamma}(t))=\bm{v}(t)+z\bm{\omega}(t)\times \bm{e}(t)$ is the rod's velocity at the collision point.
In Eq.~\eqref{eq_frequency}, the first, second, and third terms in the curly bracket are relevant to the collision on the side ($\|$) and two opposing ($\pm$) edges of the rod.
Based on $f(z,\bm{n}; \bm{\Gamma}(t))$ and $F(\bm{\Gamma})$, the coordinate of the collision point and the collision time interval $\tau$ between successive collisions are sampled using stochastic techniques \cite{Devroye1986}.
(ii) From these sampled variables, $\bm{r}$, $\bm{e}$, $\bm{v}$, and $\bm{\omega}$ are updated based on the rules of classical mechanics for a rigid body.
Repeating these samplings and updates, we calculate the dynamics of the mobile rod.
The details of the derivation of the collision statistics, sampling method, and the update scheme are described in Appendix\ \ref{appendix_kmc_method}.

\begin{figure*}[htbp]
\centering
 \includegraphics[width=0.8\textwidth]{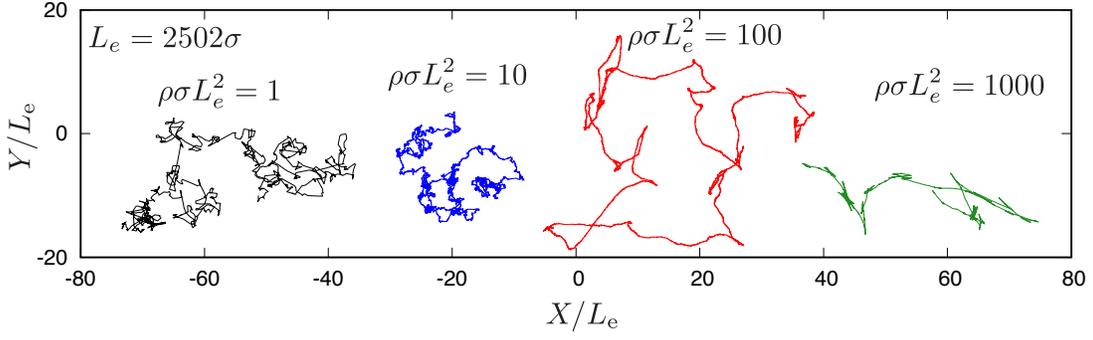}
\caption{Trajectories of the rod's ($L_e=2502\ \Revised{\sigma}$) center of mass for various scaled obstacle densities $\rho \Revised{\sigma}L_e^2$ from the KMC simulation.
3D motions are projected onto the $XY$-plane and scaled by $L_e$.
}
\label{fig_trajectory}
\end{figure*}

\section{results}
%\textit{Results. ---}
Qualitatively different behaviors occur during a change of $\rho$ at fixed \Revised{$L_e=2502\ \sigma$}, as visually captured by representative trajectories in Fig.~\figref{fig_trajectory}.
The observed time duration is \Revised{$2.0\times 10^6 \sigma/u$}.
For \Revised{$\rho \sigma L_e^2=1$} and $10$, the mobile rod seems to move randomly.
At higher number densities \Revised{$\rho \sigma L_e^2=100$}, the straight motion persists over longer distances compared with those for lower densities \Revised{$\rho \sigma L_e^2=1$} and $10$.
For \Revised{$\rho \sigma L_e^2=1000$}, we observe straight and bouncing motions.

\begin{figure}[tbp]
\centering
 \includegraphics[width=0.8\columnwidth]{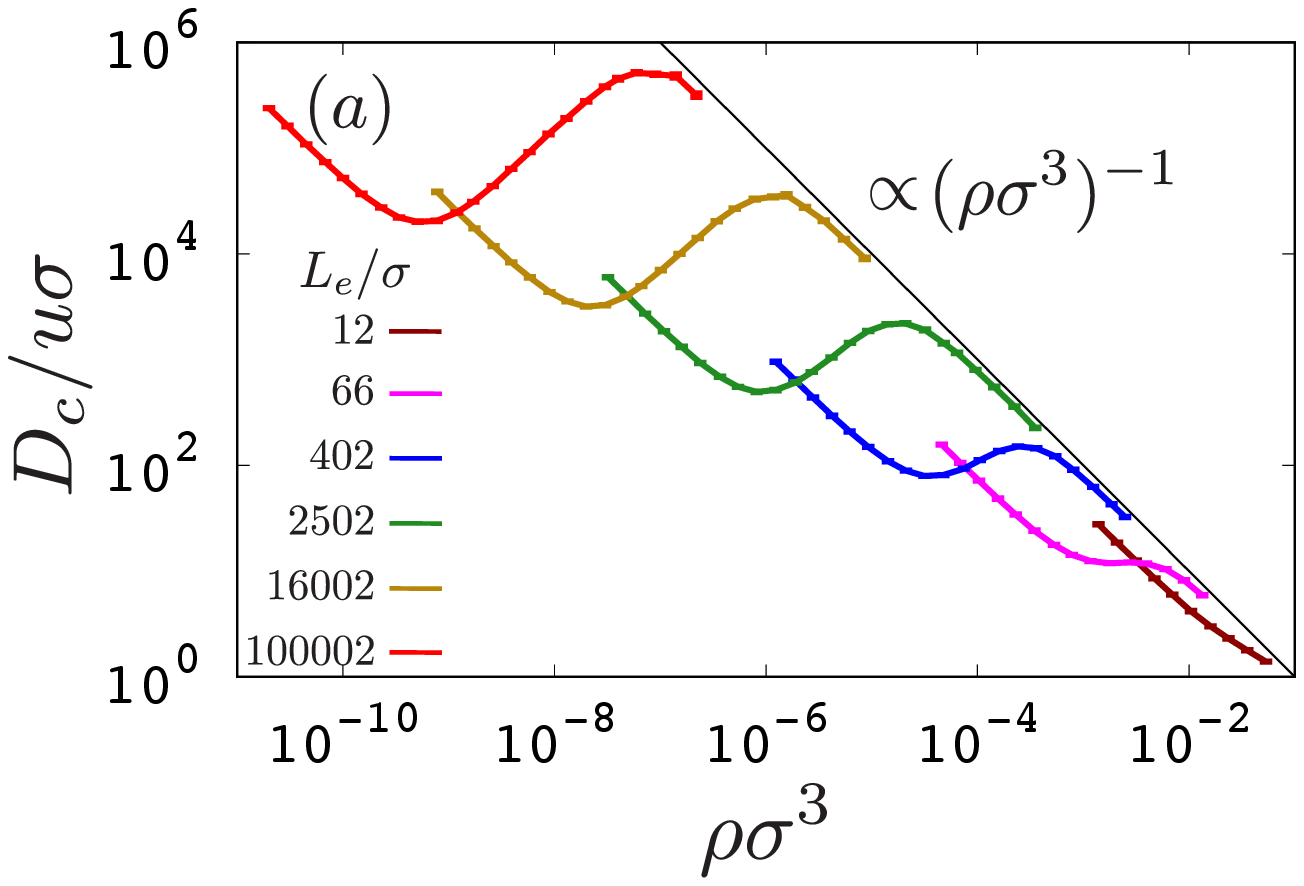}\\[1mm]
\includegraphics[width=0.81\columnwidth]{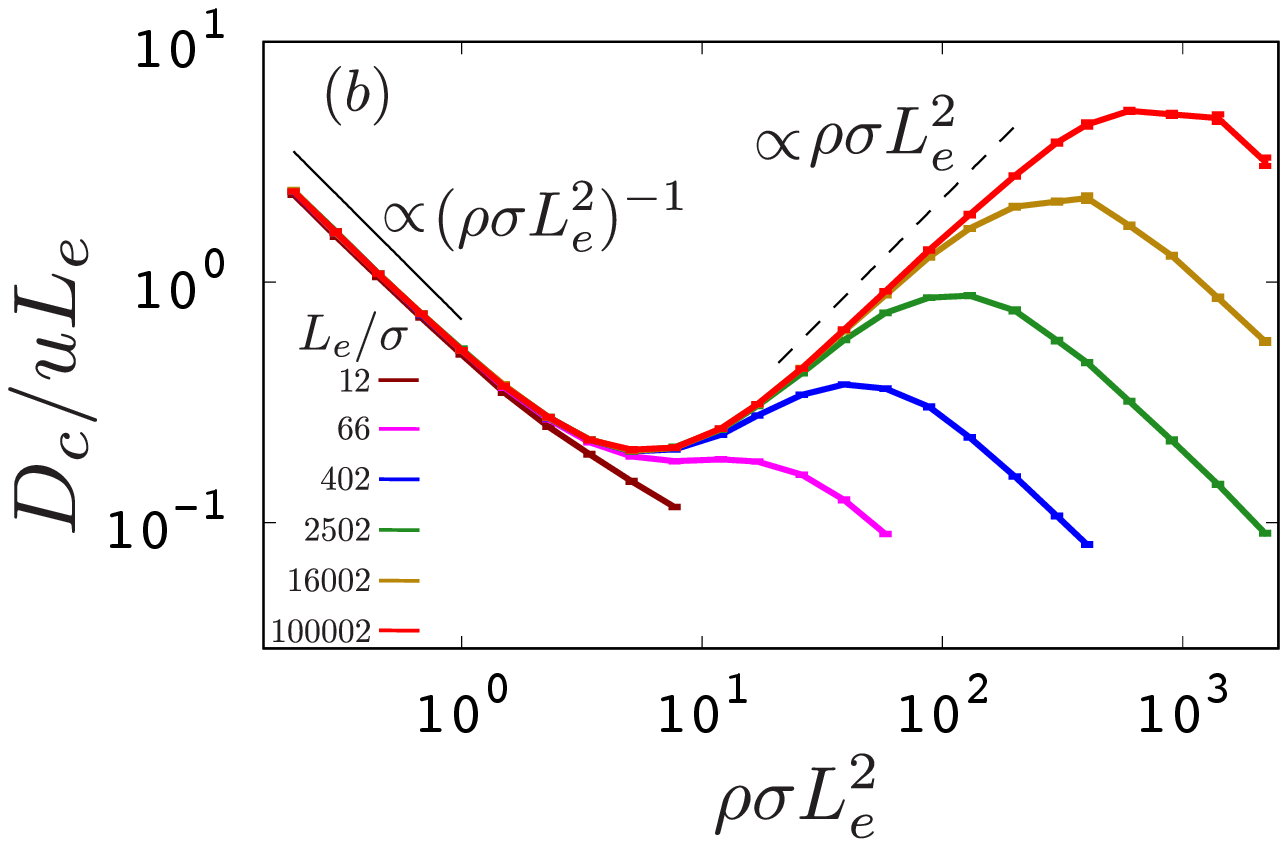}
 \caption{
 Translational diffusion coefficient of the mobile rod (various rod lengths $L_e$) from the KMC simulations.
 Data are shown as (a) $D_c\Revised{/u\sigma}$ versus $\rho\Revised{\sigma^3}$ and (b) in scaled form $D_c\Revised{/u L_e}$ versus $\rho \Revised{\sigma}L_e^2$.
 Error bars and asymptotic exponents are also displayed.
 \label{fig_dc_kmc}}
\end{figure}

To quantify these motions (Fig.~\figref{fig_trajectory}),
we calculate $D_c$ of the mobile rod from its center-of-mass mean square displacement (MSD) in the linear time domain.
\Revised{$D_c/u\sigma$} versus the obstacle number density \Revised{$\rho\sigma^3$} are displayed in Fig.~\figref{fig_dc_kmc}(a) for various mobile rod lengths \Revised{$L_e$} (error bars arise from the linear fitting).
In this figure, $D_c$ shows non-monotonic behaviors with increasing $\rho$ for the highly elongated rods \Revised{$L_e\gtrsim 66\ \sigma$}; $D_c$ at large $L_e$ exhibits both a local minimum and maximum.
When the same data are represented in scaled forms, \Revised{$D_c/u L_e$} and \Revised{$\rho\sigma L_e^2$}, as shown in Fig.~\figref{fig_dc_kmc}(b), the curves collapse except for the larger density regime.
From Figs.~\figref{fig_dc_kmc}(a,b), the asymptotic forms are observed for small, intermediate, and large density regimes as \Revised{$D_c/u\sigma\propto (\rho \sigma^2 L_{e})^{-1}$}, \Revised{$D_c/u\sigma\propto \rho L_{e}^3$}, and
\Revised{$D_c/u\sigma\propto (\rho\sigma^3)^{-1}$}, respectively.
We emphasize that the non-monotonic $\rho$ dependency for $D_c$ arises even under the Markovian process.
In contrast to $D_c$, the rotational diffusion coefficient $D_r$ in the current system exhibits monotonic behavior against $\rho$, \Revised{$D_r \sigma/u \sim (\rho L_e^3)^{-1}$} as shown in Fig.~\figref{fig:dr_kmc} (Appendix \ref{app:moredata}).

% \mkcomment{I propose to delete this paragraph below as the details can be found in the SI}
% In contrast to $D_c$, the rotational diffusion in the current system exhibits plain monotonic behavior with increasing obstacle density.
% We defined the rotational diffusion coefficient of the mobile rod $D_r$ as $D_r\sim(2\tau_r)^{-1}$, where $\tau_r$ is the rotational relaxation time. $\tau_r$ is obtained from the curve fitting for the directional correlation function $\langle \bm{e}(t)\cdot\bm{e}(0)\rangle$ via the exponential function $\exp(-t/\tau_r)$. We treat only the data in the larger density regime $\rho L_e^2>3$ since the $\langle \bm{e}(t)\cdot\bm{e}(0)\rangle$ shows damped oscillation in the smaller density regime $\rho L_e^2<3$.
% The obtained rotational diffusion coefficient converges to $D_r\sim (\rho L_e^3)^{-1}$ as shown in Fig.~S4 in the supplementary material.

The scaling relations between $D_c$ and $\rho$ are simply explained based on the Markovian nature.
Here, $D_c$ is also calculated from the integration of the velocity auto-correlation function over time lag\Revised{: $D_c=\int_0^{\infty} \langle \bm{v}(t)\cdot\bm{v}(0)\rangle dt$}, instead of the mean square displacement. Thus, the diffusion coefficient would be approximated as the \Revised{product of the square of speed $u^2$ and the relaxation time of the center of mass velocity.}
The collision frequency can be decomposed into two contributions: collision frequencies from the side $F_{\|}$ and edges $F_{\pm}$.
These contributions scale as \Revised{$F_{\|} \sim \rho \sigma L_eu$} and \Revised{$F_{\pm}\sim \rho\sigma^2u$}. These estimates are confirmed by the rigorous calculations for the collision frequencies as shown in Eqs.~\eqref{eq_frequency_side} and \eqref{eq_frequency_edge} in Appendix.
The average angular velocity scales as \Revised{$\bar{\omega}\sim u/L_e$}.
In the dilute regime \Revised{$\rho \sigma L_e^2\lesssim 1$}, the relation $\bar{\omega}>F_{\|}$ is satisfied.
In this low density regime, the rod mainly rotates and occasionally collides with an obstacle on its side. By a few collisions, the motion of the rod largely changes since the rod experiences the impulsive forces from various directions.
Then, \Revised{the relaxation time of the center of mass velocity is approximated as the mean collision time $\sim 1/F_{\|}$ and $D_c$ scales as $D_c\sim u^2/F_{\|}\sim u/\rho \sigma L_e $}.
This description is consistent with the observed random motions for the lower density regimes \Revised{$\rho \sigma L_e^2\lesssim 10$} in Fig.~\figref{fig_trajectory}.
In the higher density regime \Revised{$\rho \sigma L_e^2\gtrsim 1$}, where the relation $\bar{\omega} \Revised{<} F_{\|}$ is fulfilled, the rotational motion of the rod is diffusive, and thus the direction of the rod slowly changes.
In this density regime, the velocity with the orthogonal direction rapidly relaxes, whereas that with the axial direction is not largely disturbed.
In such a case, there are possible relaxation mechanisms for the velocity with axial direction: the change of rod direction or the collision on the edge.
Here, the change of rod direction between collisions is approximately $\Delta \theta\sim \bar{\omega}/F_{\|}$, and the rotational relaxation time scales as $\tau_{\rm{r}}\sim \Delta\theta^{-2}/F_{\|}\sim \rho L_e^3\Revised{\sigma/u}$. This estimate also predicts the rotational diffusion coefficient $D_r=(2\tau_r)^{-1}\sim \Revised{u/\rho L_e^3 \sigma}$, in full agreement with our measurements, c.f., Fig.~\figref{fig:dr_kmc} (Appendix \ref{app:moredata}).
The collision time interval on the edge is about $F_{\pm}^{-1}$.
In the intermediate density regime where $D_c$ increases, the rotational relaxation time is smaller than the collision time interval on the edge.
Thus, the velocity relaxes by the rotation of the direction, and consequently the diffusion coefficient is approximately \Revised{$D_c\sim u^2\tau_{r}\sim\rho L_e^3 u\sigma$}.
Within the high density regime where $D_c$ decreases again, the collision on the edge is the main mechanism causing velocity relaxation with the axial direction, and we obtain \Revised{$D_c\sim u^2/F_{\pm}\sim u/\rho\sigma^2$}.
These mechanisms explained above seem to be consistent with the persistence of the straight motion with \Revised{$\rho \sigma L_e^2=100$} and the straight and bouncing motions with \Revised{$\rho \sigma L_e^2=1000$} displayed in Fig.~\figref{fig_trajectory}, and the estimated exponents agree with the simulation results in Fig.~\figref{fig_dc_kmc}.

%\Revised{I'll mention about lower $L_e$ (or also characteristic $\rho$) here.}
%\mkcomment{Please also see fitting appendix} 

One may suspect that the increase in $D_c$ is an artifact since we assume a Markovian process even in the high density regime.
However, we next show that this assumption is indeed a good approximation to calculate $D_c$ for a rod embedded in a 3D sea of point obstacles.
To this end, we calculate the dynamics of a rod using conventional molecular dynamics (MD) simulations \cite{allen1989computer}.
Here, instead of a hard-core potential, the repulsive Weeks-Chandler-Andersen potential \cite{weeks1971role} is employed for the elastic interaction between rod and point obstacles.
The details of the simulation method are described in Appendix~\ref{appendix_md}.
Figure~\figref{fig_dc_md} displays \Revised{$D_c/u L_e$ (symbols) against $\rho\sigma L_e^2$ with various reduced rod lengths $L_e/\sigma$} obtained via MD. Error bars are again calculated from linear fitting for the MSDs.
Due to the computational cost, data for large rod lengths \Revised{$L_e\gtrsim16000\ \sigma$} could not be sampled.
For comparison, the KMC data from Fig.~\figref{fig_dc_kmc} are shown in Fig.~\figref{fig_dc_md} (solid curves).
The MD results quantitatively agree with those obtained via KMC.
This indicates that multi-body correlations are negligible in estimating $D_c$ within the explored wide regime of $\rho$.

\begin{figure}[tbp]
 \begin{center}
  \includegraphics[width=0.9\columnwidth]{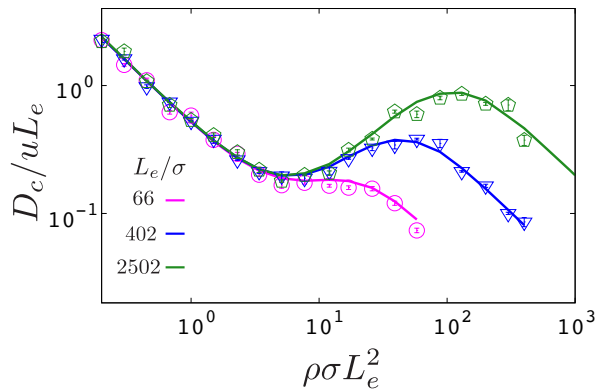}
\caption{
Reduced translational diffusion coefficient $D_c\Revised{/u L_e}$ versus obstacle density \Revised{$\rho\sigma L_e^2$} with the error bars from MD simulations (Symbols).
Data for three rod lengths $L_e$ are displayed.
For comparison, the KMC simulation results (Fig.~\figref{fig_dc_kmc}) are shown by curves.}
\label{fig_dc_md}
 \end{center}
\end{figure}

\section{Discussion}
%\textit{Discussion.---}
This work shows that $D_c$ increases even in a Markovian process and that the observed exponents are easily rationalized.
This result does not imply that the exponents in prior studied systems can be simply understood.
Frenkel and Maguire \cite{frenkel1981molecular, frenkel1983molecular} investigated $D_c$ of a constituent particle in a system of infinitely thin hard rods, where $D_c$ was found to be proportional to the root of the rod density.
For a 2D rod in the presence of point obstacles studied by H\"{o}fling, Frey, and Franosch \cite{hofling2008enhanced}, the power exponent of $D_c$ versus obstacle density is $0.8$ in the concentrated regime.
Mandal et al \cite{mandal2020crowding} investigated the dynamics of a rod-shaped active swimmer (along the axial direction) and showed that $D_c$ depends on the square of the density of the constituent.
In these prior systems, the kinetic constraints are not negligible, and they should be taken into account to explain the exponents.

Some works investigated similar systems to ours.
Tucker and Hernandez \cite{tucker2010observation, tucker2011absence} numerically studied the dynamics of a $5$\AA{} long mobile rod in the presence of spatially fixed spherical obstacles with radius $0.5$\AA{} with rod thickness $0$, $0.1$, and $0.5$\AA.
They argued that the increase in $D_c$ does not occur in their 3D system, while it can occur in the corresponding 2D setup.
One may think that these findings are inconsistent with our results.
However, if one identifies the lengths in their system with ours, the effective aspect ratio of the rod becomes about $10$ since the interaction distance between the rod and the obstacle is the rod thickness plus obstacle size.
For the rod with such an aspect ratio $10$, an increase in $D_c$ does not occur\Revised{, as shown in Appendix\ \ref{appendix_fitting}; the upturn of $D_c$ emerges for 
an aspect ratio $\gtrsim 24$  corresponding to $L_e/\sigma\gtrsim 520/11$ in our setting.}
Conversely, in Tucker and Hernandez's system, the increase in $D_c$ will occur for a much smaller obstacle radius or much larger rod length.
Otto, Aspelmeier, and Zippelius \cite{otto2006microscopic} theoretically analyzed the dynamics of a constituent particle of infinitely thin rods under the assumption of the Markovian process\Revised{; the sequential collisions are uncorrelated.}
%They argued that an increase in $D_c$ should not occur under such circumstances.
\Revised{They argued that an increase in $D_c$ should not occur under the Markovian process.}
This result obviously contradicts our findings.
However, they did not consider the long-time persistence of the ballistic motion with the axial direction. Thus, the increase in $D_c$ could not be captured \Revised{in their theory}.

It should be emphasized that a rise of $D_c$ can occur for a ballistic system \cite{frenkel1981molecular, frenkel1983molecular, magda1986transport, magda1988transport, hofling2008enhanced} or some active matter systems \cite{mandal2020crowding} due to the persistence of the motion with the axial direction.
One may think that an increase in $D_c$ can occur for passive rod-shaped particles in some solvents or porous media.
However, it can not exhibit the increase in diffusivity by the same mechanism as our system since the persistence of the motion with axial direction rapidly relaxes by the Brownian motion.
Recently, the increase in diffusivity with increasing aspect ratio is observed for rod in a gel \cite{rose2022shape}, although the mechanism would be different to our system.

The current system consists of a rod colliding with immobile, or infinitely heavy point obstacles. Let us consider the situation where obstacles move in an equilibrium state. As long as the obstacle mass is sufficiently larger than $M$, the obstacle motion is slow because of the Maxwell-Boltzmann velocity distribution. In this case, the situation would not be largely different from the current system since the moving particles can be approximated as the fixed obstacles for the rod particle, and the increase in diffusivity will emerge.
In contrast, if the obstacle mass is comparable to $M$, the situation can be different from the current system since the translational and rotational relaxation times vary largely with the obstacle mass. Even in this case, the increase in the diffusivity can emerge since it simply originates from the reduction of the rotational motion and the persistence of the axial motion.
The analyses for the effects of obstacle mass on the increase in diffusivity will be interesting work.

\section{Conclusions} 

% This study demonstrated that a $D_c$ upturn can emerge even in Markovian nature, where the kinetic constraint does not exist.
% As a simple model system, we investigated the single mobile rod-shaped particle in immobile fixed obstacles in three-dimensions using highly efficient Markovian kinetic Monte Carlo simulations.
% The translational diffusion coefficient of the rod decreases, increases, and decreases again as the obstacle density increases.
% These non-trivial behaviors could be explained based on the Markovian process.
% This work sheds light on the kinetics of non-spherical-shaped particles.

This study demonstrated that a $D_c$ upturn can emerge even in Markovian nature, where the kinetic constraint does not exist.
As a simple model system, we investigated the single mobile rod-shaped particle in immobile fixed obstacles in three-dimensions using highly efficient kinetic Monte Carlo simulations under the Markovian process.
The translational diffusion coefficient of the rod decreases, increases, and decreases again as the obstacle density increases.
This non-trivial diffusivity could be explained based on the Markovian process.
\Revised{Namely, without using the concept of the kinetic constraint, the upturn of $D_c$ was simply explained via two time-related quantities: the angular velocity and the collision frequency.
Further, the lower limit of $L_e$, where the upturn of $D_c$ emerges in our system, was estimeted as $L_e/\sigma \simeq 520/11$ (aspect ratio $24$).}
This work sheds light on the kinetics of non-spherical-shaped particles.

%\acknowledgements
\begin{acknowledgments}
FN and MK were supported by the “Young Researchers Exchange Programme between Japan and Switzerland” under the “Japanese-Swiss Science and Technology Programme”.
FN was also supported by Grant-in-Aid for JSPS (Japan Society for the Promotion of Science) Fellows (Grant No.
JP21J21725).
\end{acknowledgments}

\appendix
\setcounter{figure}{5}

\section{Kinetic Monte Carlo Method}\label{appendix_kmc_method}
\subsection{Overview}
For the calculation of the stochastic dynamics, the kinetic Monte Carlo (KMC) method is employed.
Classical KMC has been originally used for chemical reactions \cite{gillespie1976general} the Ising model \cite{bortz1975new} and lateron for many other systems \cite{chatterjee2007}.
In this work, we extend the conventional KMC method to calculate the dynamics of our rod-shaped particle. To perform the KMC simulation, we impose the following four assumptions:
(i) The point obstacles are homogeneously distributed in three-dimensional space,
(ii) The dynamics of the mobile rod is a Markovian process,
(iii) The interaction between the mobile rod and the point obstacles is hard-core potential; the point obstacles do not penetrate into the mobile rod, and
(iv) The collision frequency is constant during a small time duration $\Delta t$. 

 %
%\begin{itemize}
% \item[\smallblacksquare] The point obstacles are homogeneously distributed in three-dimensional space.
% \item[\smallblacksquare] The dynamics of the mobile rod is a Markovian process.
% \item[\smallblacksquare] The interaction between the mobile rod and the point obstacles is hard-core potential; the %point obstacles do not penetrate into the mobile rod.
% \item[\smallblacksquare] The collision frequency is constant during a small time duration $\Delta t$. 
%\end{itemize}
From these, we calculate the collision frequency for a given rod direction, velocity and angular velocity, as described below.
Based on this collision frequency, the collision time interval and the coordinate of the collision point on the rod surface are sampled. With this information at hand, the position, velocity, direction, and angular velocity of the mobile rod are updated.
For a rod-shaped particle with a transient symmetry axis ${\bf e}(t)$, the collision frequency is not a constant between successive collisions since it depends on $\bm{e}(t)$. The collision time interval distribution is therefore not a simple exponential function, and the time interval cannot be straightforwardly sampled. The fourth assumption above has been made to avoid this difficulty.
Due to this assumption, the current implementation involves an additional parameter $\Delta t$.
We checked that the obtained KMC results are insensitive to the precise value of $\Delta t$ provided it is small compared with the mean rotation period of the rod as shown in Fig.~\figref{fig:dc_kmc_dt01}.
All \Revised{KMC simulation} results to be presented in the manuscript have been obtained using $\Delta t$ small compared with the mean rotation period, more precisely, using $\Delta t = \sqrt{I/\kBT}/100$, where $I$ is the relevant component of the moment of inertia tensor, and $\kBT/2$ the mean kinetic energy for one degree of freedom.

\subsection{Collision frequencies}
Here we derive the collision statistics between the rod and the point obstacles.
The rod length, radius, mass, and inertia tensor are denoted by $L$, $\sigma$, $M$, $\bm{I}$, 
while the total kinetic energy is $5\kBT/2$.
The number density of the point obstacle is $\rho$.
%In the main manuscript, we choose $\sigma$, $M$, and $\kBT$ to define the dimensionless units, however, the following quantities are displayed with dimensions for physical clarity.
\Revised{As in the main manuscript, the physical quantities are displayed with dimensions for physical clarity.}
The rod's center of mass velocity, its angular velocity, and the direction of its symmetry axis at time $t$ are denoted as $\bm{v}(t)$, $\bm{\omega}(t)$, and $\bm{e}(t)$, and the following relation is always satisfied,
\begin{equation}
 \bm{e}(t)\cdot\bm{\omega}(t)=0.
\end{equation}
because $\dot{\bf e}=\bm{\omega}\times{\bf e}$ and because the length of ${\bf e}$ does not change in time, $0={\bf e}\cdot\dot{\bf e}$, where the dot denotes a derivative with respect to time $t$.
In the following, we denote by $\bm{\Gamma}(t)$ the 8-dimensional time-dependent phase space variable $({\bf v}(t), \bm{\omega}(t),{\bf e}(t))$ characterizing
the state of the rod and furthermore introduce the effective velocity on the rod surface as $\bm{v}_e(z;\bm{\Gamma}(t))=\bm{v}(t)+z\bm{\omega}(t)\times \bm{e}(t)$, where $z$ is the axial coordinate along $\bm{e}(t)$ in the range $z\in[-L/2,L/2]$.

A collision with a point obstacle occurs either at the cylindrical surface (side) or at the surface of one of the two half-spherical bases (edges) of the rod-like particle. The collision point on the surface is specified by $z$ and the direction of the surface normal $\bm{n}$ at the collision point (a unit vector) as shown in Fig.~\figref{fig:rod_variables}.
If the collision occurs at the side, $\bm{n}$ is perpendicular to $\bm{e}$(t).
For the collision occurring at the edges, $z$ is $+L/2$ or $-L/2$.
From these, the collision frequency density at $z$ and $\bm{n}$ for a given $\bm{\Gamma}(t)$, $f(z,\bm{n}; \bm{\Gamma}(t))$ can be expressed as the sum over three terms corresponding to the side and the two edges, 
\begin{equation}
  f(z,\bm{n}; \bm{\Gamma}(t))
  = \sum_{\mu\in\{\|,-,+\}} f_\mu(z,\bm{n}; \bm{\Gamma}(t)) 
\end{equation}
%\begin{equation}
%  f(z,\bm{n}; \bm{\Gamma}(t))
%  =f_\|(z,\bm{n}; \bm{\Gamma}(t)) + f_+(z,\bm{n}; \bm{\Gamma}(t))
%  + f_-(z,\bm{n}; \bm{\Gamma}(t))
%\end{equation}
with 
\begin{eqnarray}
f_{\|}(z,\bm{n}; \bm{\Gamma}(t)) 
&=&\rho\sigma\bm{v}_e(z;\bm{\Gamma}(t))\cdot\bm{n}
\Theta[\bm{v}_e(z;\bm{\Gamma}(t))\cdot\bm{n}]
\times \nonumber \\
&& \delta(\bm{e}(t)\cdot\bm{n}),
\label{eq_frequency_z_n_side} 
\end{eqnarray} 
and
\begin{eqnarray} 
f_{\pm}(z,\bm{n}; \bm{\Gamma}(t)) 
&=&\rho\sigma^2\bm{v}_e(z;\bm{\Gamma}(t))\cdot\bm{n}\
\Theta[\bm{v}_e(z;\bm{\Gamma}(t))\cdot\bm{n}]
\times
\nonumber \\ && \Theta[\pm\bm{e}(t)\cdot\bm{n}]\delta\left(z\mp \frac{L}{2}\right).
\label{eq_frequency_z_n_pm}
\end{eqnarray}
Here, $\Theta$ is the Heaviside step function, and $\delta$ is the Dirac delta distribution.
To obtain the collision frequency, we need to integrate the density $f(z,\bm{n}; \bm{\Gamma}(t))$ over the surface of the rod-like particle, i.e., over $z$ and $\bm{n}$. To this end it is convenient to 
choose two suitable, and therefore different, frames of reference for the two types of contributions
given by Eqs.~\eqref{eq_frequency_z_n_side} and \eqref{eq_frequency_z_n_pm}.

% \begin{figure}
%     \centering
%     \includegraphics[width=0.6\columnwidth]{fig_appendix_system_kmc.eps}
%     \caption{Schematic figure of the rod. The collision frequency for a given $\bm{e}(t)$, $\bm{v}(t)$, and $\bm{\omega}(t)$ is characterized by $z$ and $\bm{n}$.}
%     \label{fig:rod_variables}
% \end{figure}

\paragraph{Integration of $f_\|$.} 
%$\blacksquare$ 
For the integration of Eq.~\eqref{eq_frequency_z_n_side}, we set the frame of reference so that the direction vector of the rod and the effective velocity become $\bm{e}(t)=(0, 0, 1)$ and $\bm{v}_e(t)=(v_r, 0, v_z)$ with $v_r>0$.
Then, the direction vector at the collision point can be described as $\bm{n} = (\sin\theta\cos\phi, \sin\theta\sin\phi, \cos\theta)$ with $\theta\in [0, \pi]$ and $\phi\in [-\pi, \pi]$.
In this coordinate system, the integration of Eq.~\eqref{eq_frequency_z_n_side} over $\theta$ yields
\begin{eqnarray}
    f_\|(z,\phi; \bm{\Gamma}(t)) 
&=&\int_0^{\pi} d\theta
f_\|(z,\theta, \phi; \bm{\Gamma}(t))
\nonumber \\
&=&\rho \sigma v_r(z;\bm{\Gamma}(t))\cos(\phi)\
\Theta[\cos\phi],
\label{eq_frequency_z_phi_side}
\end{eqnarray}
and the subsequent integration over the remaining angle $\phi$ of the surface normal leads to the collision frequency density on the side at $z$ 
\begin{equation}
f_\|(z; \bm{\Gamma}(t)) 
    =\int_{-\pi}^{\pi} d\phi
    f_\|(z,\phi; \bm{\Gamma}(t)) 
=2\rho \sigma v_r(z;\bm{\Gamma}(t)).
\label{eq_frequency_z_side}
\end{equation}
From Eq.~\eqref{eq_frequency_z_side}, the total collision frequency at the side $F_\|(\bm{\Gamma}(t))$ evaluates to 
\begin{eqnarray}
    F_\|(\bm{\Gamma}(t)) 
  &=& \int_{-L/2}^{L/2}dz\ f_\|(z; \bm{\Gamma}(t)) \label{eq_frequency_side} \\
  &=&\frac{\rho \sigma L^2 \omega(t)}{4}
\left[
(\alpha+1)c_+-(\alpha-1)c_-
-\chi 
\right], \nonumber
\end{eqnarray}
where we made use of the abbreviations $\omega(t)=|\bm{\omega}(t)|$, $v(t)=|\bm{v}(t)|$, 
$c_\pm  = \sqrt{(\alpha\pm 1)^2+\beta}$, and
 \begin{eqnarray}
    \alpha &=& \frac{2\bm{v}(t)\cdot(\bm{\omega}(t)\times \bm{e}(t))}{\omega^2(t) L}, \nonumber \\
  \beta  &=& \frac{4(\bm{v}(t)\cdot\bm{\omega}(t))^2}{\omega^4(t) L^2}, \nonumber \\
  \chi &=& \beta\ln\left(\frac{\alpha-1+c_-}{\alpha+1+c_+}\right).
 \end{eqnarray}
For $\bm{v}\|\bm{e}$, this reduces to $\alpha=\beta=0$ and thus $F_\|(\bm{\Gamma}(t))=\rho \sigma L^2\omega/2$, which vanishes for $\omega=0$, as expected.

\paragraph{Integration of $f_\pm$.} 
%$\blacksquare$ 
For the integration of Eq.~\eqref{eq_frequency_z_n_pm}, we employ the new frame of reference such that $\bm{v}_{e}(z; \bm{\Gamma}(t))=(0,0,v_e(z;\bm{\Gamma}(t)))$ defines the new $z$-axis, and $\bm{e}(t)=(e_r, 0, e_z)$ with $e_r>0$ fixes the new $x$-axis. Then, the surface normal is described as $\bm{n} = (\sin\theta\cos\phi, \sin\theta\sin\phi, \cos\theta)$ with $\theta\in [0, \pi]$ and $\phi\in [-\pi, \pi]$.
In this coordinate system, collision frequency density at the angle vector $\bm{n}$ is simply given by
\begin{eqnarray}
    f_{\pm}(\bm{n}; \bm{\Gamma}(t)) 
&=&\int_{-L/2}^{L/2}dz f_{\pm}(z,\bm{n}; \bm{\Gamma}(t)) \nonumber \\ 
&=&\rho \sigma^2 \bm{v}_e(\pm L/2;\bm{\Gamma}(t))\cdot \bm{n}
\Theta[\pm\bm{e}(t)\cdot\bm{n}] \times 
\nonumber \\
&&\Theta[\bm{v}_e(\pm L/2;\bm{\Gamma}(t))\cdot \bm{n}].
\label{eq_frequency_n_pm}
\end{eqnarray}
By next integrating Eq.~\eqref{eq_frequency_n_pm} over $\phi$, we obtain the collision frequency density at $\theta$ as
\begin{eqnarray}
%\begin{split}
f_{\pm}(\theta; \bm{\Gamma}(t)) 
&=&\int_{-\pi}^{\pi} d\phi f_{\pm}(\bm{n}; \bm{\Gamma}(t)) 
 \\
&=&2\rho \sigma^2 v_e(\pm L/2;\bm{\Gamma}(t))\cos(\theta)\ \Theta[\cos\theta]\times  \nonumber \\
&&\hspace*{-8mm}\bigl\{
\pi\Theta\left[\pm\gamma(\theta)\!-\!1\right]
+\cos^{-1}\!\left(\mp\gamma(\theta)\right)
\Theta\left[1\!-\!\left|\gamma(\theta)\right|\right]
\bigr\}, \nonumber 
%\end{split}
\end{eqnarray}
where we defined $\gamma(\theta)=e_z\cos\theta/e_r\sin\theta$.
The collision frequencies at the two edges are obtained by performing the remaining integration over $\theta$, 
\begin{eqnarray}
    F_{\pm}(\bm{\Gamma}(t)) 
    &=&\int_0^{\pi} \sin\theta d\theta\, f_{\pm}(\theta; \bm{\Gamma}(t))
    \label{eq_frequency_edge} %\\
    %&=&
    = 
    \frac{\pi\rho\sigma^2}{2}
    \times \\ &&
    \bigl\{\ |\bm{v}_e(\pm L/2;\bm{\Gamma}(t))|
    \pm \bm{v}_e(\pm L/2;\bm{\Gamma}(t))\cdot \bm{e}(t)\bigr\}.
     \nonumber
\end{eqnarray}

%$\blacksquare$ 
\paragraph{Combining the integrated $f_\|$ and $f_\pm$.}
Combining Eqs.~\eqref{eq_frequency_side} and \eqref{eq_frequency_edge}, the total collision frequency for given $\bm{v}(t)$, $\bm{\omega}(t)$, and $\bm{e}(t)$ becomes
\begin{equation}
    F(\bm{\Gamma}(t)) = F_\|(\bm{\Gamma}(t)) 
    + F_{+}(\bm{\Gamma}(t)) + F_{-}(\bm{\Gamma}(t)).
    \label{eq_frequency_total}
\end{equation}

\subsection{Time evolution}
To calculate the trajectory of the mobile rod, we require the collision time interval $\tau$ between successive collisions and information about the collision point at the surface of the rod specified by $z$ and $\bm{n}$.
Making use of the fourth assumption, the probability density $P(\tau)$ on the side and the edges for a given 
phase space coordinate $\bm{\Gamma}(t)$ can be described using Eq.~\ref{eq_frequency_total} as
\begin{equation}
    P(\tau)=F(\bm{\Gamma}(t))\exp[-F(\bm{\Gamma}(t)) \tau],
    \label{eq_prob_total}
\end{equation}
where this probability density is reasonable in $0<\tau<\Delta t$. % , where we set $\Delta t=\sqrt{I/\kBT}/100$.

%$\blacksquare$ 
(i) Based on Eq.~\eqref{eq_prob_total}, $\tau$ is sampled using inversion method with the range $\tau\in[0, \infty]$.

%$\blacksquare$ 
(ii) If $\tau>\Delta t$, a collision does not occur in $\Delta t$.
In this case, the time $t$, position $\bm{r}(t)$, and direction vector $\bm{e}(t)$ are updated to $t+\Delta t$, $\bm{r}(t+\Delta t)=\bm{r}(t)+\bm{v}(t)\Delta t$, and $\bm{e}(t+\Delta t)=(0, -\sin(\omega \Delta t), \cos(\omega\Delta t))$ in the frame of reference where $\bm{e}(t)=(0,0,1)$ and $\bm{\omega}(t)=(\omega,0,0)$ define the $z$- and $x$-axes,
while $\bm{v}$ and $\bm{\omega}$ remain unchanged as $\bm{v}(t+\Delta t)=\bm{v}(t)$ and $\bm{\omega}(t+\Delta t)=\bm{\omega}(t)$.
After that, we sample $\tau$ again for the updated variables, $\bm{\Gamma}(t+\Delta t)$.

%$\blacksquare$ 
(iii) If $\tau<\Delta t$, the time $t$, position $\bm{r}(t)$, and direction $\bm{e}(t)$ are similarly updated to $t+\tau$, $\bm{r}(t+\tau)=\bm{r}(t)+\bm{v}(t)\tau$, and $\bm{e}(t+\tau)=(0, -\sin\omega \tau, \cos\omega\tau)$.
Using new $\bm{\Gamma}(t+\tau)$, we determine whether the collision occurs at the side or the edges based on the collision frequencies given by Eqs~\eqref{eq_frequency_side} and \eqref{eq_frequency_edge}; the probabilities where the collision occurs at the side is $F_\|(\bm{\Gamma}(t+\tau))/F(\bm{\Gamma}(t+\tau))$ and those at the edges are $F_{\rm{\pm}}(\bm{\Gamma}(t+\tau))/F(\bm{\Gamma}(t+\tau))$.

%\smallblacksquare 
(iiia) When the collision occurs at the side, $z$ is sampled from the probability density $P(z; \bm{\Gamma}(t+\tau))$ obtained from Eqs.~\eqref{eq_frequency_z_side} and \eqref{eq_frequency_side} as
\begin{eqnarray}
     P(z; \bm{\Gamma}(t+\tau))
 &=& \frac{f_\|(z; \bm{\Gamma}(t+\tau))}{F_\|(\bm{\Gamma}(t+\tau))}
=\frac{v_r}{\mathcal{N}}
 \\
&=&\frac{|\left[\bm{1}-\bm{e}(t+\tau)\bm{e}(t+\tau)\right]\bm{v}_e(z;\bm{\Gamma}(t+\tau))|}{\mathcal{N}},
\nonumber
\end{eqnarray}
where $\bm{1}$ is the unit tensor, and $\mathcal{N}$ is the appropriate normalization factor.
From this probability density, we sample $z$ in the range $z\in[-L/2, L/2]$ using the rejection method.
The probability density of $\phi$ is obtained from Eqs.~\eqref{eq_frequency_z_side} and \eqref{eq_frequency_z_phi_side} as
\begin{equation}
 P(\phi; \bm{\Gamma}(t+\tau))
 =\frac{ f_\|(z, \phi;  \bm{\Gamma}(t+\tau))}{ f_\|(z ;  \bm{\Gamma}(t+\tau)} = \frac{\cos(\phi)\, \Theta[\cos\phi]}{2}.
\end{equation}
We sample $\phi$ using the inversion method via $\phi=\sin^{-1}(1-2\xi)$ with the equally distributed random numbers $\xi\in[0,1]$.
For the collision at the side, $\theta$ is trivially $\pi/2$.
Thus, the collision point $z$ and $\bm{n}$ at time $t+\tau$ for a given $\bm{\Gamma}(t+\tau)$ are determined.

%\smallblacksquare 
(iiib) When the collision occurs at one of the edges at $z=\pm L/2$, the probability density for $\bm{n}$ is constructed from Eq.~\eqref{eq_frequency_n_pm} and \eqref{eq_frequency_edge} as
\begin{eqnarray}
    P(\bm{n}; \bm{\Gamma}(t+\tau))
    &=&\frac{f_{\pm}(\bm{n}; \bm{\Gamma}(t+\tau))\sin\theta}
    {F_{\pm}(\bm{\Gamma}(t+\tau))}
     \\
    &=&\frac{\cos(\theta)\sin(\theta)\
    \Theta[\cos\theta]\
    \Theta[\pm\bm{e}(t+\tau)\cdot\bm{n}]}{\mathcal{N}},
    \nonumber
    \label{eq_prob_n_edge}
\end{eqnarray}
where $\mathcal{N}$ is another normalization factor, and the additional term $\sin(\theta)$ is from the solid angle.
From Eq.~\eqref{eq_prob_n_edge}, $\theta$ and $\phi$ are simultaneously sampled by the rejection method.
Then, $z$ and $\bm{n}$ characterizing the collision point at time $t+\tau$ for a given $\bm{\Gamma}(t+\tau)$ are determined.

%\smallblacksquare 
(iv) The velocity and angular velocity just after a collision, $\bm{v}(t+\tau)$ and $\bm{\omega}(t+\tau)$ are updated based on sampled $z$ and $\bm{n}$.
In this calculation, we employ the frame of reference where $\bm{e}(t+\tau)=(0,0,1)$ defines $z$-axis.
In this frame, the two components in inertia tensor $I_{xx}$ and $I_{yy}$ are $I$ as
\begin{equation}
    I_{xx}=I_{yy}=I=\Revised{M}\frac{5L^3+20\sigma L^2+45\sigma^2L+32\sigma^3}{60L+80\sigma}.
    \label{eq_inertia_moment}
\end{equation}
Here, the hard-core potential between sphero-cylinder and point obstacles does not make the torque along $z$-axis.
Thus, the rod does not rotate around $z$-axis, and we need not consider $I_{zz}$.
The non-diagonal components of $\bm{I}$ in this frame are zero because of the symmetry of the rod shape.
$\bm{v}(t+\tau)$ and $\bm{\omega}(t+\tau)$ are calculated as follows:
\begin{align}
  \bm{v}(t+\tau) &= \bm{v}(t)+\Delta v\ \bm{n}  \label{updatev} \\
 I\bm{\omega}(t+\tau) &= I\bm{\omega}(t)+ \Delta v\ z\ \bm{e}(t+\tau)\times \bm{n}, \label{updateomega}
\end{align}
where $\Delta v$ is the magnitude of the velocity change. It is obtained via energy conservation before and after the collision, 
$ v^2(t)+I\omega^2(t)=v^2(t+\tau)+I\omega^2(t+\tau)$
as
\begin{equation}
 \Delta v = -\frac{2I \left[\bm{v}(t)+z\bm{\omega}(t)\times\bm{e}(t+\tau)
\right]\cdot\bm{n}}
{I+z^2(\bm{e}(t+\tau)\times\bm{n})^2}.
\end{equation}
%$\blacksquare$ 
(v) After the update for $\bm{v}(t+\tau)$ and $\bm{\omega}(t+\tau)$, we sample $\tau$ again, i.e., start over at (i). We repeat this algorithm to calculate the time series of the position, velocity, direction, and angular velocity of the mobile rod.

\begin{figure}[tbp]
 \begin{center}
  \includegraphics[width=0.8\columnwidth]{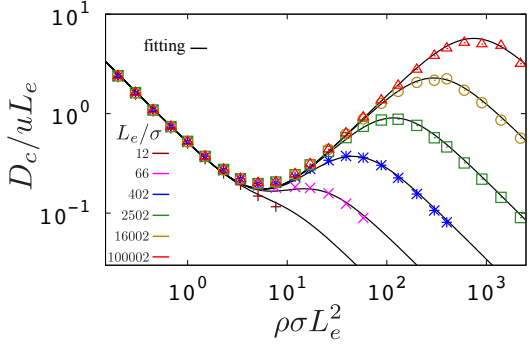}
\caption{\Revised{Fitting function for $D_c$, Eq.~\eqref{fit}, with various rod lengths $L_e$ (black curves).
For comparison, $D_c$ from the KMC simulation in Fig.\ \figref{fig_dc_kmc}(b) are also displayed by symbols. According to Eq.~\eqref{fit}, the upturn is estimated to be absent for $L_e/\sigma \le 520/11$.}
}
\label{fig_kmc_fitting}
 \end{center}
\end{figure}

\Revised{
\section{Analytic expression for $D_c$} \label{appendix_fitting}
A both qualitatively and quantitatively correct fitting function of our measured $D_c$ data is
\begin{equation}
    \frac{D_c}{u L_e} \approx \frac{1}{2\rho\sigma L_e^2}+\frac{11\rho \sigma L_e^2}{65[11+2(\rho\sigma L_e^2)^2\sigma/L_e]}, \label{fit}
\end{equation}
which by construction satisfies the three scaling relations (1) $D_c/u\sigma \sim 1/\rho\sigma^2L_e$, 
(2) $D_c/u\sigma \sim \rho L_e^3$, and (3) $D_c /u\sigma \sim 1/\rho\sigma^3$ observed in Fig.\ \figref{fig_dc_kmc} at the small, intermediate and large $\rho$ limits, respectively.
We can also confirm that Eq.~\eqref{fit} agrees quantitatively with the KMC simulation data in Fig.~\figref{fig_dc_kmc}(b), as displayed in Fig.~\figref{fig_kmc_fitting}.
From this expression \eqref{fit}, $D_c/u L_e$ monotonically decreases with increasing $\rho\sigma L_e^2$ when $L_e/\sigma\le 520/11$ (aspect ratio is about $24$), otherwise $D_c/u L_e$ has a local minimum and maximum against increasing $\rho \sigma L_e^2$.
Also from \eqref{fit}, one can estimate the two crossover densities $\rho_{1\leftrightarrow 2}$ and $\rho_{2\leftrightarrow 3}$ 
as follows: $\rho_{1\leftrightarrow 2} \approx 5.7\times (\sigma L_e^2)^{-1}$ and $\rho_{2\leftrightarrow 3} \approx 2.3\times (\sigma L_e)^{-3/2}$.
They are both independent of temperature, while $D_c$ increases linearly with $u=\sqrt{\kBT}$ in fixed obstacle geometry
(the relation $D_c\propto u$ is generally observed in gas diffusion \cite{dorfman2021, erpenbeck1991self, dymond1985hard})}.

%\mkcomment{Please check, I added something. Attention, We have incompatible scalings in the low density regime in the manuscript. The one I am stating here is from the data you sent for Fig 4, but above we state $(\rho L_e^3)^{-1}$. Is it not surprising that $D_c$ is proportional to $\nu$, i.e., $\sqrt{\kBT}$? Is the maximal possible $\rho_{\rm max}\approx (\pi \sigma^2 L_e)^{-1}$? Then we may convert the crossover densities to crossover volume fractions to eventually come up with limiting $L_e$? for example $\rho_{1\leftrightarrow 2} \ll \rho_{\rm max}$ implies $L_e \gg 5.7\ \sigma$?}
%\fmcomment{I was wrong for $D_c$ in the small density regime and revised it. $D_c\propto \sqrt{\kBT}$ wouldn't be surprising because it is often found in gas diffusion. Concerning $L_e$ that shows $D_c$ upturn, I wrote a note on file "calc\_lower\_L.tex" in the folder note on overleaf, and such a $L_e$ is about $47\sigma$.}

\section{Additional simulation data from KMC simulations}\label{app:moredata}
\begin{figure}[tbp]
    \centering
    \includegraphics[width=0.8\columnwidth]{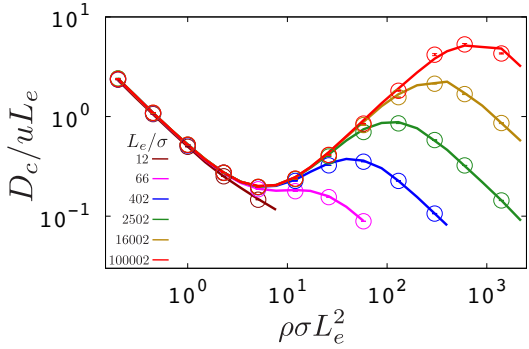}
    \caption{\Revised{Reduced} translational diffusion coefficient $D_{c}\Revised{/u L_e}$ versus the scaled obstacle density $\rho \Revised{\sigma}L_e^2$ for various rod lengths $L_e$, from the KMC simulations. Symbols represent the data with $\Delta t=\sqrt{I\Revised{/\kBT}}/10$. For comparison, the data with $\Delta t=\sqrt{I\Revised{/\kBT}}/100$ in the main manuscript are also displayed by curves.}
    \label{fig:dc_kmc_dt01}
\end{figure}

\begin{figure}[tbp]
    \centering
    \includegraphics[width=0.8\columnwidth]{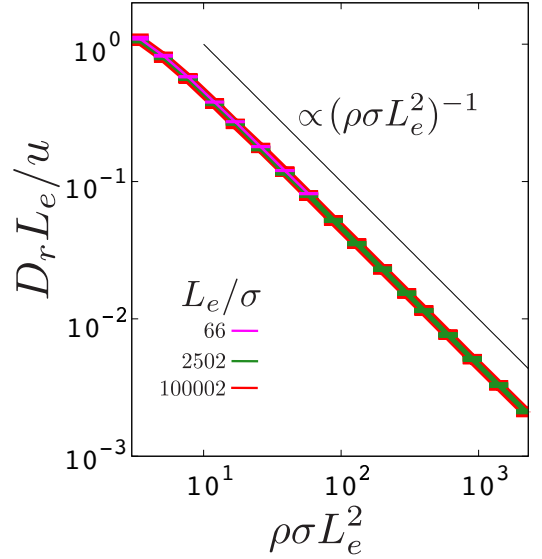}
    \caption{Reduced rotational diffusion coefficient $D_rL_e\Revised{/u}$ against the scaled obstacle density $\rho \Revised{\sigma}L_e^2$ from the KMC simulations. The asymptotic exponent \Revised{$D_r\sim u/\rho L_e^3 \sigma$} and error bars from the curve fittings are also displayed.
    }
    \label{fig:dr_kmc}
\end{figure}

%The following data are displayed with dimensionless quantities; we choose $\sigma$, $M$, and $\kBT$ to define the dimensionless units as for the main manuscript.
\Revised{Figure}~\figref{fig:dc_kmc_dt01} displays \Revised{$D_c/u\sigma$} against \Revised{$\rho \sigma L_e^2$} with \Revised{$\Delta t=\sqrt{I/\kBT}/10$} (symbols) and \Revised{$\Delta t=\sqrt{I/\kBT}/100$} (curves).
The data for \Revised{the two different $\Delta t$ values clearly overlap}. This result indicates that the KMC simulation result is \Revised{almost unaffected} by the size of $\Delta t$ when $\Delta t$ is sufficiently small compared with the rotational time period\Revised{: $\Delta t\ll \sqrt{I/k_BT}$}.
The scaled rotational diffusion coefficient \Revised{$D_r L_e/u$} versus the scaled obstacle density \Revised{$\rho \sigma L_e^2$} from the KMC simulations within \Revised{$\rho\sigma^2 L_e<1$} is shown in Fig.~\figref{fig:dr_kmc}. Here, $D_r$ is defined by the rotational relaxation time $\tau_r$ as $D_r=(2\tau_r)^{-1}$, where $\tau_r$ is obtained from the curve fitting for the directional correlation function $\langle \bm{e}(t)\cdot\bm{e}(0)\rangle$ using the exponential function $\exp(-t/\tau_r)$.
In this fitting, we treat the data only in the higher density regime \Revised{$\rho \sigma L_e^2>3$} where $\langle \bm{e}(t)\cdot\bm{e}(0)\rangle$ monotonically decreases and can be reasonably fit by the exponential function (in the smaller density regime \Revised{$\rho \sigma L_e^2<3$}, $\langle \bm{e}(t)\cdot\bm{e}(0)\rangle$ exhibits a damped oscillation, and it is no longer the \Revised{mono-exponentially decreasing}). 

\section{Molecular dynamics simulations}\label{appendix_md}
This system consists of a single mobile rod and, as opposed to the KMC method, a large finite number ($10^8$) of spatially fixed point obstacles at number density $\rho$ in a three-dimensional cubic box with periodic boundary conditions.
The positions of the fixed obstacles are uniformly distributed in the cubic box, and the mobile rod is initially placed without overlap to the obstacles.
The shape of the mobile rod is sphero-cylindrical, and its radius, length of axis, mass are $\sigma$, $L$, $M$, as before (Fig.\ \figref{fig:rod_variables}).
The moment of inertia tensor is $\bm{I}$ given by Eq.~\eqref{eq_inertia_moment} in the frame of reference where $\bm{e}=(0,0,1)$ defines the $z$-axis.
The total energy of the system is denoted by $E$, now with contributions from the kinetic energy for the translational and rotational degrees of freedom of the mobile rod as well as of the potential energy between the mobile rod and fixed obstacles as 
%\fmcomment{Thanks. $U$ is good. I ran the simulation with E=5$\epsilon$/2. In the simulated obstacle density, $\kBT$ is within $0.97<\kBT<1$. It means that the rod moves ballistically in almost time, while a small part of the energy goes to the potential part.}
\begin{equation}
 E = \frac{Mv^2}{2}+\frac{I\omega^2}{2}+U(\bm{r}, \bm{e}, \{\bm{R}_i\}),
\end{equation}
where $I$ is given in Eq.~\eqref{eq_inertia_moment}, 
$\bm{v}$, $\bm{\omega}$, $\bm{r}$, and $\bm{e}$ are the center of mass velocity, angular velocity, center of mass position, and direction unit vector of the mobile rod, respectively, and $\bm{R}_i$ denotes the position of the $i$-th obstacle.
In the molecular dynamics simulation, we choose $U$ as a Weeks-Chandler-Andersen (WCA) potential which works for the minimum distance between the major axis of the rod and the $i$-th obstacles $d_i=d(\bm{r}, \bm{e}, \bm{R}_i)$ as
\begin{equation}
  U(\bm{r}, \bm{e}, \{\bm{R}_i\})\\
= 4\epsilon\sum_i 
\left[\left(\frac{\sigma}{d_i}\right)^{12}-
\left(\frac{\sigma}{d_i}\right)^6+\frac{1}{4}\right],
\end{equation}
for $d_i \le 2^{1/6}\sigma$, 
where $\epsilon$ is the energy unit for the WCA potential, and $d_i=d(\bm{r}, \bm{e}, \bm{R}_i)$ is calculated using the algorithm by Lumelsky \cite{lumelsky1985fast}.
%In the molecular dynamics simulation, $M$, $\sigma$, and $\epsilon$ are chosen to define dimensionless units.
The initial $\bm{v}$ and $\bm{\omega}$ are randomly given so that the sum of $Mv^2/2$ and $I\omega^2/2$ becomes $5\epsilon/2$.
Consequently, $E$ keeps $5\epsilon/2$ because the system is prepared without overlap ($U=0$) at startup.
We calculate the dynamics of the rod using the Leap-Frog algorithm extended to the rod-shaped particle \cite{allen1989computer}.
The results are obtained for the time duration $10^7\sigma\sqrt{M/\epsilon}$ with a step size $10^{-3}\sigma\sqrt{M/\epsilon}$.
%\mkcomment{add units for time and step size}
%In Fig.~\figref{fig_dc_md}, the data are shown to define the dimensionless units by $M$, $\sigma$, and $\kBT$ to compare that from the KMC simulation.
\Revised{Fig.~\figref{fig_dc_md} displays the reduced diffusion coefficient $D_c/u L_e$ versus number density $\rho\sigma L_e^2$ for various rod lengths $L_e$.
We recall, $u=\sqrt{\kBT/M}$, and $\kBT/2$ was calculated as the mean of the kinetic energy of a single degree of freedom in the MD simulation.
}
In the molecular dynamics simulation, the rod is completely trapped at rather highly number densities $\rho \sigma^2 L_e\gtrsim1$, where $L_e=L+2\sigma$.
Such a trapping motion is beyond the scope of this work.
Thus we limit ourselves to the range of the number densities as $\rho \sigma^2 L_e<1$, or equivalently, $\rho\sigma L_e^2\le L_e/\sigma$.

\begin{figure}[tbh]
    \centering
    \includegraphics[width=0.8\columnwidth]{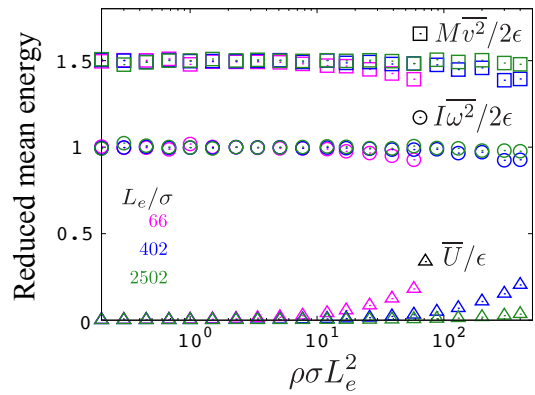}
    \caption{Reduced mean energies versus obstacle density from the MD simulations. The symbols square, circle, and triangle represent the mean translational ($M\overline{v^2}/2$), rotational ($I\overline{\omega^2}/2$), and potential ($\overline{U}$) energies, divided by $\epsilon$.
    Their sum is strictly conserved. The colors pink, blue, and green indicate the data with the rod lengths $L_e=66\ \sigma$, $L_e=402\ \sigma$, and $L_e=2502\ \sigma$, respectively. 
    %\mkcomment{Can you replace the y axis label by "Reduced mean energy"?}
    }
    \label{fig:energy}
\end{figure}

In the MD simulation, the total energy $E$ is distributed over the translational, rotational, and potential energies. We display the time-averaged energies $M\overline{v^2}/2$, $I\overline{\omega^2}/2$, and $\overline{U}$ versus the obstacle density in Fig.~\figref{fig:energy}.
From these data, $M\overline{v^2}/2\epsilon$ and $I\overline{\omega^2}/2\epsilon$ become almost $3/2$ and $1$ in the simulated density regime, although these kinetic energies slightly decrease when the obstacle density approaches the trapping density regime $\rho \sigma^2 L_e\sim 1$. 

\bibliographystyle{apsrev4-2}
\bibliography{references}

\end{document}